\documentclass[aps,twocolumn,showpacs,superscriptaddress,pre]{revtex4}
\usepackage{amsmath}
\usepackage{amssymb}
\usepackage{mathtext}
\usepackage{graphicx}
\usepackage{mathtext}

\begin{document}

\title[Near-surface bubbly horizon in porous media \& Temperature wave]{Formation of bubbly horizon in liquid-saturated porous medium by surface temperature oscillation}

\author{Denis S.\ Goldobin}
\affiliation{Institute of Continuous Media Mechanics, UB RAS,
         1 Academik Korolev str., Perm 614013, Russia}
\affiliation{Department of Theoretical Physics,
         Perm State University, 15 Bukireva str., 614990, Perm, Russia}
\affiliation{Department of Mathematics, University of Leicester,
         University Road, Leicester LE1 7RH, UK}
\author{Pavel V.\ Krauzin}
\affiliation{Institute of Continuous Media Mechanics, UB RAS,
         1 Academik Korolev str., Perm 614013, Russia}
\date{\today}

\begin{abstract}
We study non-isothermal diffusion transport of a weakly-soluble substance in a liquid-saturated porous medium being in contact with the reservoir of this substance. The surface temperature of the porous medium half-space oscillates in time, which results in a decaying solubility wave propagating deep into the porous medium. In such a system, the zones of saturated solution and non-dissolved phase coexist with the zones of undersaturated solution. The effect is firstly considered for the case of annual oscillation of the surface temperature of water-saturated ground being in contact with atmosphere. We reveal the phenomenon of formation of a near-surface bubbly horizon due to the temperature oscillation. An analytical theory of the phenomenon is developed. Further, the treatment is extended to the case of higher frequency oscillations and case of weakly-soluble solids and liquids.

 % end of PACS codes

\end{abstract}

\maketitle

\section{Introduction}\label{sec1}
Diffusion transport in bubbly media~\cite{Donaldson-etal-1997-1998,Haacke-Westbrook-Riley-2008,Goldobin-Brilliantov-2011,Krauzin-Goldobin-2014} and media with condensed non-dissolved phase~\cite{Haacke-Westbrook-Riley-2008,Davie-Buffett-2001,Goldobin-CRM-2013,Goldobin-etal-EPJE-2014} appears to possess unique features for isothermal conditions and is even more intriguing for non-isothermal ones. For these systems, the non-dissolved phase makes the local solute concentration being equal to the solubility. The solute concentration is not a free variable any more but a function of temperature and pressure. Simultaneously, the non-zero divergence of the solute flux determined by the solubility gradient does not change the enslaved solute concentration but redistributes the mass of non-dissolved phase. Thus, for the dynamics of systems with non-dissolved phase, novel phenomena and mechanisms, which never appear for undersaturated solutions, come into play. These phenomena are especially strongly pronounced for systems where the non-dissolved phase is immobilised ({\it e.g.}, in porous media) and solubility is small~\cite{Goldobin-Brilliantov-2011}. For an immobilised non-dissolved phase, the mass transport operates only through the solution, and when the solubility is small the mass accumulated in the non-dissolved phase can be several orders of magnitude larger than the dissolved mass.

In Ref.\,\cite{Krauzin-Goldobin-2014}, we reported the effect of surface temperature oscillations, which produce the solubility wave, on the diffusive transport in porous media where the non-dissolved phase is present everywhere and discussed the physical systems for which the effect is relevant. The systems where the zones with non-dissolved phase can coexist with the zones of undersaturated solution can exhibit richer and more sophisticated dynamics. A liquid-saturated porous medium being in contact with the reservoir of weakly-soluble substance ({e.g.}, atmosphere) is a system of that kind. In this paper we consider the effect of the surface temperature oscillation on diffusive transport in the porous medium half-space being in contact with the reservoir of a weakly-soluble substance.

For the sake of definiteness, we will consider transport of gases in detail and comment how the results can be extended to the case of weakly-soluble solids ({\it e.g.}, limestone) and liquids ({\it e.g.}, crude oil). The case of solids and liquids is different from the case of gases only by insensitivity of the solubility to pressure. From the view point of mathematics, for the processes under our consideration, the case of gas hydrates is also identical to the case of weakly-soluble solids.

Mathematically, the contact with atmosphere is represented via the boundary condition; at the contact boundary the solute concentration equals the solubility at any instant of time. The same boundary condition and, consequently, the phenomena we report in this paper will hold for the case of contact with reservoir of any substance.

In this paper, we will reveal that the temperature wave leads to formation of a near-surface bubbly horizon and `oversaturation' of the medium with the atmosphere gas compared to the period-mean solubility. In particular, the net molar fraction (non-dissolved phase + solution) of the gas molecules in pores next to the surface equals the maximal-over-period solubility.

The phenomenon we will report is common and can be important for various systems with different origins of the surface temperature oscillations, including technological systems (filters, porous bodies of nuclear and chemical reactors, etc.). However, for the sake of convenience, we first focus on the case characterised by the hydrostatic pressure gradient which is significant for geological systems, where pressure doubles on the depth of $10$ meters, leading to a significant change of solubility. The no-pressure-gradient case will be considered as well.

The effect of enhanced filling of water-saturated ground with atmospheric gases creates more favorable conditions for local microflora and fauna and influences conditions for geochemical processes. With weakly-soluble solids, the effect provides opportunities for filling the porous matrix with some weakly-soluble ``guest'' substance in technology; the spatial mass distribution pattern of the ``guest'' substance can be controlled by the surface temperature waveform. For natural methane hydrate deposits in seafloor sediments, the effect of temperature waves on the deposit and the gas release from it is of interest in relation to the natural Glacial-Interglacial cycles~\cite{iceage} and potential global climate change~\cite{Hunter-etal-2013}.

The paper is organised as follows. In Sec.\,\ref{sec21}, we introduce the physical model for diffusion transport of weakly-soluble substance in the presence of non-dissolved phase and governing equations. With the results of numerical simulation of governing equations for a single-gas-component atmosphere, we demonstrate the phenomenon of formation of a bubbly horizon in Sec.\,\ref{sec22}. The analytical theory of the phenomenon is developed in Sec.\,\ref{sec23}. In Sec.\,\ref{sec24}, we report the dependence of integral quantifiers of the bubbly horizon on the control parameters of the system. In Sec.\,\ref{sec3}, the case of no effect of the hydrostatic pressure gradient is studied (the case corresponds to high-frequency temperature oscillations or solutions of condensed phases). The results are summarised in Conclusion (Sec.\,\ref{sec4}).

\section{Diffusion in saturated and undersaturated solutions}
\label{sec2}
\subsection{Physical model and governing equations}
\label{sec21}
The diffusion transport of solute in the presence of non-dissolved phase is essentially controlled by solubility. For moderate pressure and far from the solvent boiling point the solubility of gas in liquid reads~\cite{Pierotti-1976}
\begin{equation}
 X^{(0)}(T,P) \simeq
 X^{(0)}(T_0,P_0)\frac{T_0}{T}\frac{P}{P_0}
 \exp\left[q\left(\frac{1}{T}-\frac{1}{T_0}\right)\right]\,,
\label{eq-21-01}
\end{equation}
where molar solubility $X^{(0)}$ is the molar amount of solute per 1 mole of solvent, $T_0$ and $P_0$ are reference values, the choice of which is guided merely by convenience reason, and $X^{(0)}(T_0,P_0)$ is the solubility at the reference temperature and pressure; the parameter $q\equiv-G_i/k_\mathrm{B}$, with $G_i$
being the interaction energy between a solute molecule and the surrounding solvent molecules and $k_\mathrm{B}$ being the Boltzmann constant, is provided in table~\ref{params} for several typical gases. For condensed matter (solids and liquids) the solubility is nearly independent of pressure and approximately reads
\begin{equation}
 X_\mathrm{cond}^{(0)}(T,P) \simeq
 X_\mathrm{cond}^{(0)}(T_0,P_0)
 \exp\left[q\left(\frac{1}{T}-\frac{1}{T_0}\right)\right]\,.
\label{eq-21-02}
\end{equation}

Geological systems are typically much more uniform in the horizontal directions than in the vertical one. Hence, it is reasonable to restrict our consideration to the one-dimensional case; the system is assumed to be homogeneous in the horizontal directions. We assume the $z$-axis to be oriented downwards and its origin to be on the porous medium surface.

Let us consider harmonic oscillation of the surface temperature, $T_0+\Theta_0\cos{\omega t}$, where $T_0$ is the mean temperature, $\Theta_0$ is the oscillation amplitude, $\omega$ is the temperature oscillation cyclic frequency. In particular, annual oscillations of surface temperature only slightly deviate from their harmonic reduction (e.g., see~\cite{Yershov-1998}). The heat diffusion equation
 $\partial T/\partial t=\chi\mathrm{\Delta}T$
with no-heat-flux condition deep below the surface (at infinity) and imposed surface temperature yields
\begin{equation}
T(z,t)=T_0+\Theta_0e^{-kz}\cos(\omega t-kz)\,,
\quad
k=\sqrt{\omega/2\chi}\,,
\label{eq-21-03}
\end{equation}
where $\chi$ is the heat diffusivity and $z$ is the distance from the surface of porous medium. The hydrostatic pressure field reads
\begin{equation}
P=P_0+\rho gz\,,
\label{eq-21-04}
\end{equation}
where $P_0$ is the atmospheric pressure, $\rho$ is the liquid density, and $g$ is the gravity.

%%%%%%%%%%%%%%%%%%%%%%%%%%%%%%%%%%%%%%%%%%%%%%%%%%%%%%%%%%%%%%%%%%%%%%%
\begin{table}[t]
\caption{Chemical physical properties of solutions of nitrogen, oxygen, methane and carbon dioxide in water. Eq.\,(\ref{eq-21-01}) with $q$ and $X^{(0)}(T_0,P_0)$ specified in the table fits the experimental data from~\cite{solubility}. Eq.\,(\ref{eq-21-10}) with provided values of effective radius $R_d$ and parameter $\nu$ of the solute molecules fits the experimental data from~\cite{diffusion}.}
\begin{center}
\begin{tabular}{cp{0.7cm}p{0.7cm}p{0.7cm}p{0.55cm}}
\hline\hline
 & $\mathrm{N_2}$ & $\mathrm{O_2}$ & $\mathrm{CH_4}$ & $\mathrm{CO_2}$
 \\
\hline
$q=-G_i/k_\mathrm{B}$ (K)
 & 781 & 831 & 1138 & 1850 \\[5pt]
$X^{(0)}(20^\circ\mathrm{C},1\,\mathrm{atm})$ ($10^{-5}$)
 & 1.20 & 2.41 & 2.60 & 68.7 \\[5pt]
$R_d$ ($10^{-10}\,\mathrm{m}$)
 & 1.48 & 1.29 & 1.91 & 1.57 \\[5pt]
$\nu$ ($10^{-5}\,\mathrm{Pa\cdot s}$)
 & 9.79 & 16.3 & 28.3 & 4.68 \\[3pt]
\hline\hline
\end{tabular}
\end{center}
\label{params}
\end{table}
%%%%%%%%%%%%%%%%%%%%%%%%%%%%%%%%%%%%%%%%%%%%%%%%%%%%%%%%%%%%%%%%%%%%%%%

Since the non-dissolved phase is immobilised in pores, the mass transport in the system is contributed solely by the diffusion through the intersticial liquid and governed by equation
\begin{equation}
\frac{\partial X_\Sigma}{\partial t}
 =\nabla\cdot\left[DX_s\left(
\frac{\nabla X_s}{X_s}+\alpha\frac{\nabla T}{T}\right)\right],
\label{eq-21-05}
\end{equation}
where $X_s$ is the molar concentration of the solution, $X_\Sigma=X_s+X_b$ is the net molar fraction of gas molecules in the intersticial fluid, $X_b$ is the molar fraction of the gaseous phase (bubbles) in the intersticial fluid, $D$ is the effective molecular diffusion coefficient and $\alpha$ is the thermodiffusion constant~\cite{Bird-Stewart-Lightfoot-2007}. As compared to the molecular diffusion coefficient in bulk of pure liquid, say $D_\mathrm{mol}$, the effective coefficient is influenced by the pore network geometry (tortuosity) and the adsorption of the diffusing agent on porous matrix (on the time scales of our interest the adsorption does not lead to anomalous diffusion; it only changes the effective rate of normal diffusion~\cite{Gregg-Sing-1982}). The importance of
thermal diffusion was demonstrated for
gases~\cite{Goldobin-Brilliantov-2011} and methane
hydrate~\cite{Goldobin-CRM-2013,Goldobin-etal-EPJE-2014} on
geological time scales, although for the system of our interest it can be neglected~\cite{Krauzin-Goldobin-2014}. Henceforth, we omit the thermal diffusion term. The solute concentration
\begin{equation}
X_s=\min\{X^{(0)},\,X_\Sigma\}\,,
\label{eq-21-06}
\end{equation}
{\it i.e.}, it equals the solubility $X^{(0)}$ where the net amount of gas molecules $X_\Sigma$ exceeds the solubility, and equals $X_\Sigma$, otherwise. In the latter case, $X_b=0$.

The formulated mathematical model of the system implies that the dissolution process (as well as opposite process of formation of the non-dissolved phase from solution) occurs much faster than the change of the temperature field and the diffusive redistribution of solute mass on the macroscopic scales. In real systems, the dissolution time scales even for solid non-dissolved phase are assessed as hours (see~\cite{Buffett-Zatsepina-2000}), which is small compared to the reference times of temperature oscillation and diffusive transport on the scale of the system. The hysteresis effects possible for some phase transformations in narrow pore channels~\cite{Anderson-Tohidi-Webber-2000} are neglected in our study.

Notice, Eq.\,(\ref{eq-21-05}) is accurate for the case where macroscopic porosity is spatially uniform and the non-dissolved phase occupies a negligible fraction of the pore volume, which holds true for gases and weakly-soluble solids and liquids.

For one-dimensional case, Eq.\,(\ref{eq-21-05}) takes the form
\begin{equation}
\frac{\partial X_\Sigma}{\partial t}
 =\frac{\partial}{\partial z}\left[D\frac{\partial X_s}{\partial z}\right].
\label{eq-21-07}
\end{equation}
At the upper boundary we assume a contact with the atmosphere, which means
\begin{equation}
X_s(z=0,t)=X^{(0)}\big(T(z=0,t),P_0\big)\,.
\label{eq-21-08}
\end{equation}
Deep below the surface we assume the no-flux condition and the absence of the non-dissolved phase;
\begin{equation}
\left.\frac{\partial X_s}{\partial z}\right|_{z=+\infty}=0\,,
\qquad
X_b(z=+\infty)=0\,.
\label{eq-21-09}
\end{equation}
Notice, two boundary conditions are required at $z\to+\infty$; however, due to specificity of our system, one boundary condition, Eq.\,(\ref{eq-21-08}), is sufficient at $z=0$. Indeed, since $X_s(z=0)$ is never less than $X^{(0)}(z=0)$, the value of $X_b$ at the point $z=0$ does not influence the system dynamics; the condition for it is redundant.

Generally, all material properties of the system depend on temperature and pressure. However, feasible relative variations of the absolute temperature are small. Hence, one can neglect variation of those parameters which depend on temperature polynomially and consider variation of only those parameters which depend on temperature exponentially: the latter parameters are solubility (\ref{eq-21-01}) and the molecular diffusion coefficient $D$. Moreover, the only parameter sensitive to pressure is the gas solubility.

We employ the following dependence of molecular diffusion on temperature~\cite{Bird-Stewart-Lightfoot-2007};
\begin{equation}
D_\mathrm{mol}(T)=\frac{k_\mathrm{B}T}{2\pi\mu R_d}\cdot
 \frac{\mu+\nu}{2\mu+3\nu}\,,
\label{eq-21-10}
\end{equation}
where $\mu$ is the dynamic viscosity of the solvent, $R_d$ is the effective radius of the solute molecules with the ``coefficient of sliding friction'' $\beta$, $\nu=R_d\beta/3$. The dependence of dynamic viscosity on temperature can be described by a modified Frenkel formula~\cite{Frenkel-1955}
\begin{equation}
\mu(T)=\mu_0\exp\frac{a}{T+\tau}\,.
\label{eq-21-11}
\end{equation}
For water, coefficient
 $\mu_0=2.42\cdot10^{-5}\,\mathrm{Pa\cdot s}$,
 $a=W/k_\mathrm{B}=570\,\mathrm{K}$
($W$ is activation energy) and $\tau=-140\,\mathrm{K}$. For the effective diffusion coefficient $D$ we assume the same relative variation with temperature as for $D_\mathrm{mol}$.

%%%%%%%%%%%%%%%%%%%%%%%%%%%%%%%%%%%%%%%%%%%%%%%%%%%%%%%%%
\begin{figure*}[!t]
\centerline{
{\sf (a)}\hspace{-12pt}
\includegraphics[width=0.32\textwidth]%
 {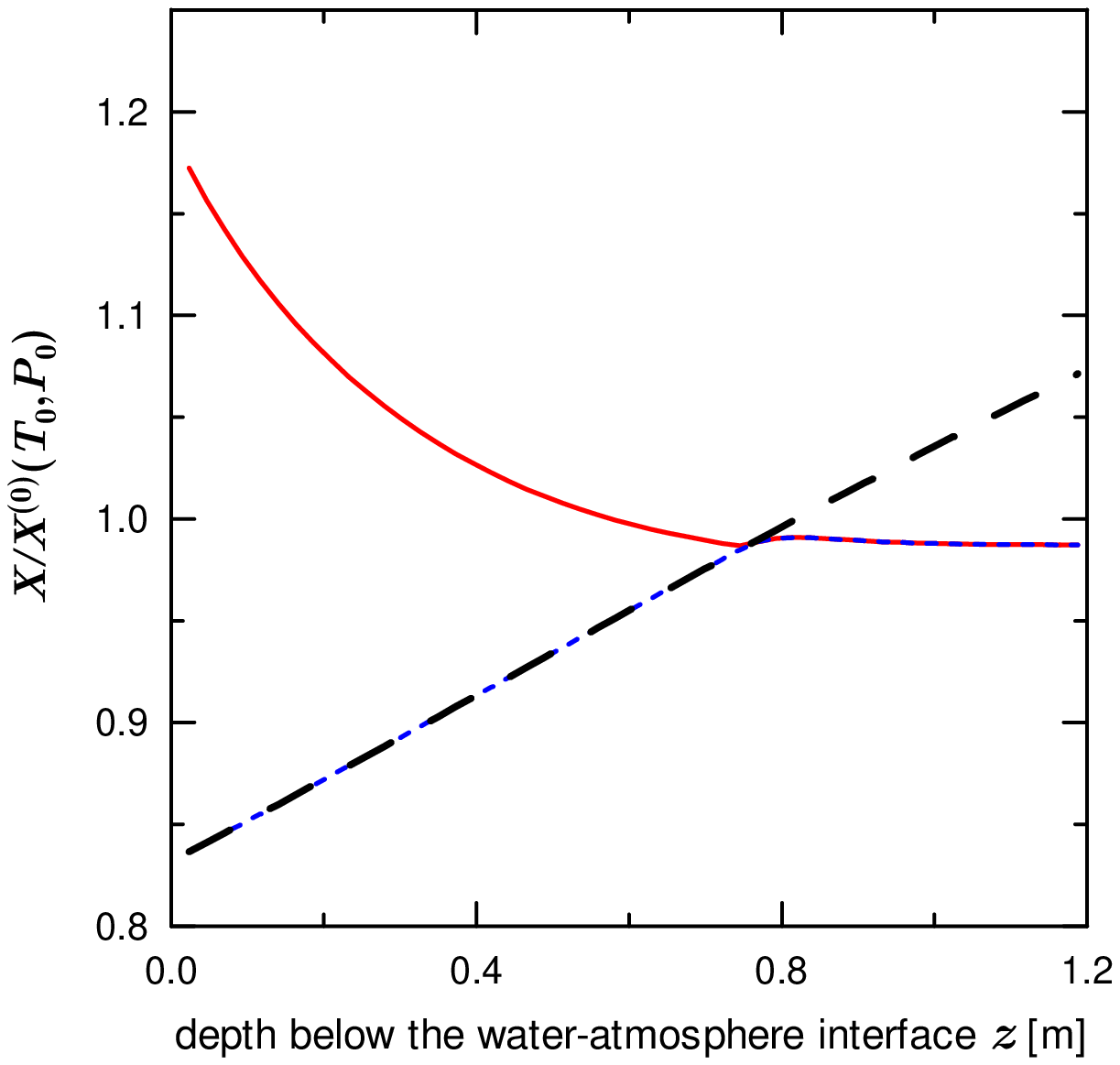}
\qquad\qquad
{\sf (b)}\hspace{-12pt}
\includegraphics[width=0.32\textwidth]%
 {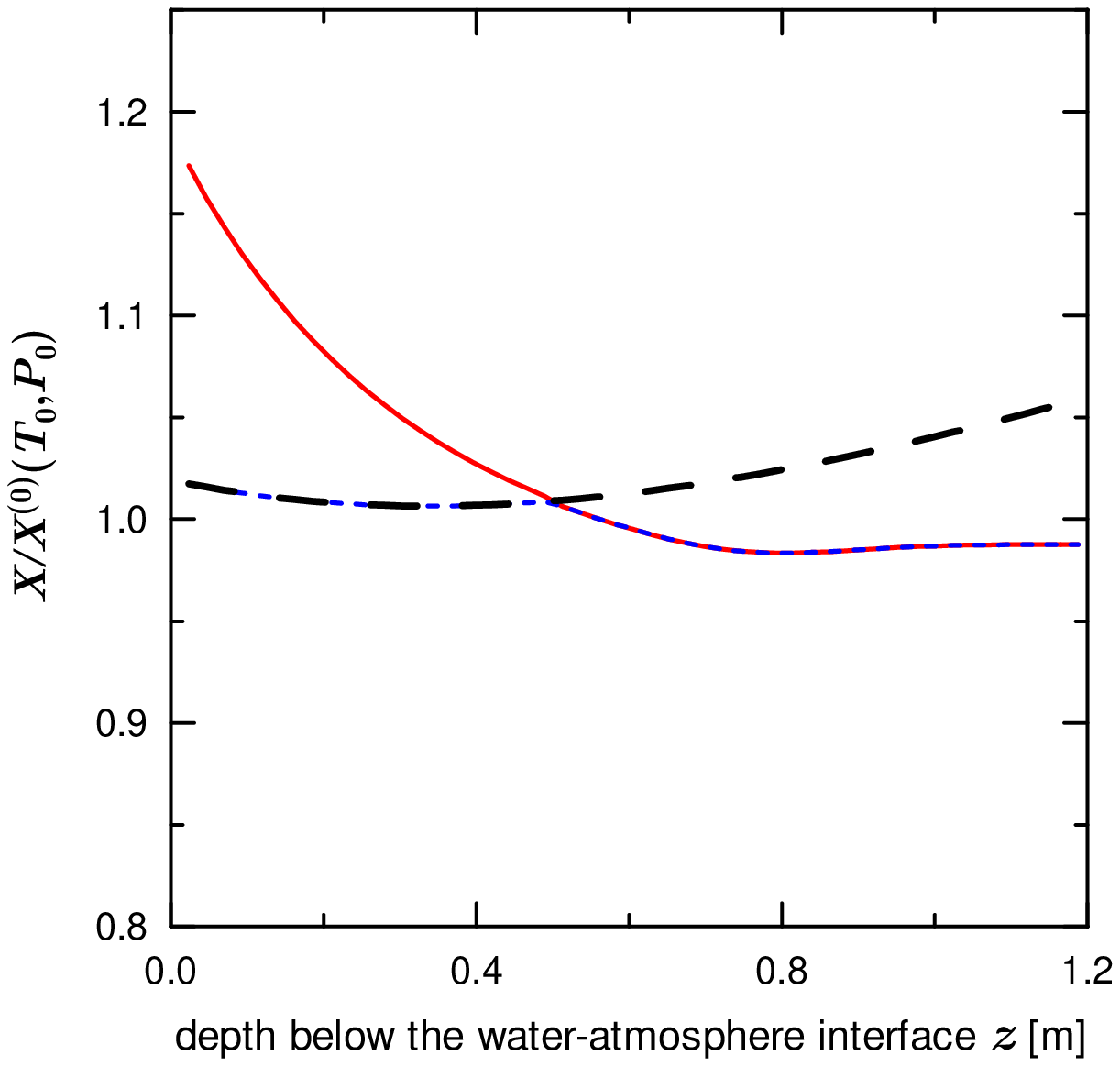}
}
\vspace{20pt}

\centerline{
{\sf (c)}\hspace{-12pt}
\includegraphics[width=0.32\textwidth]%
 {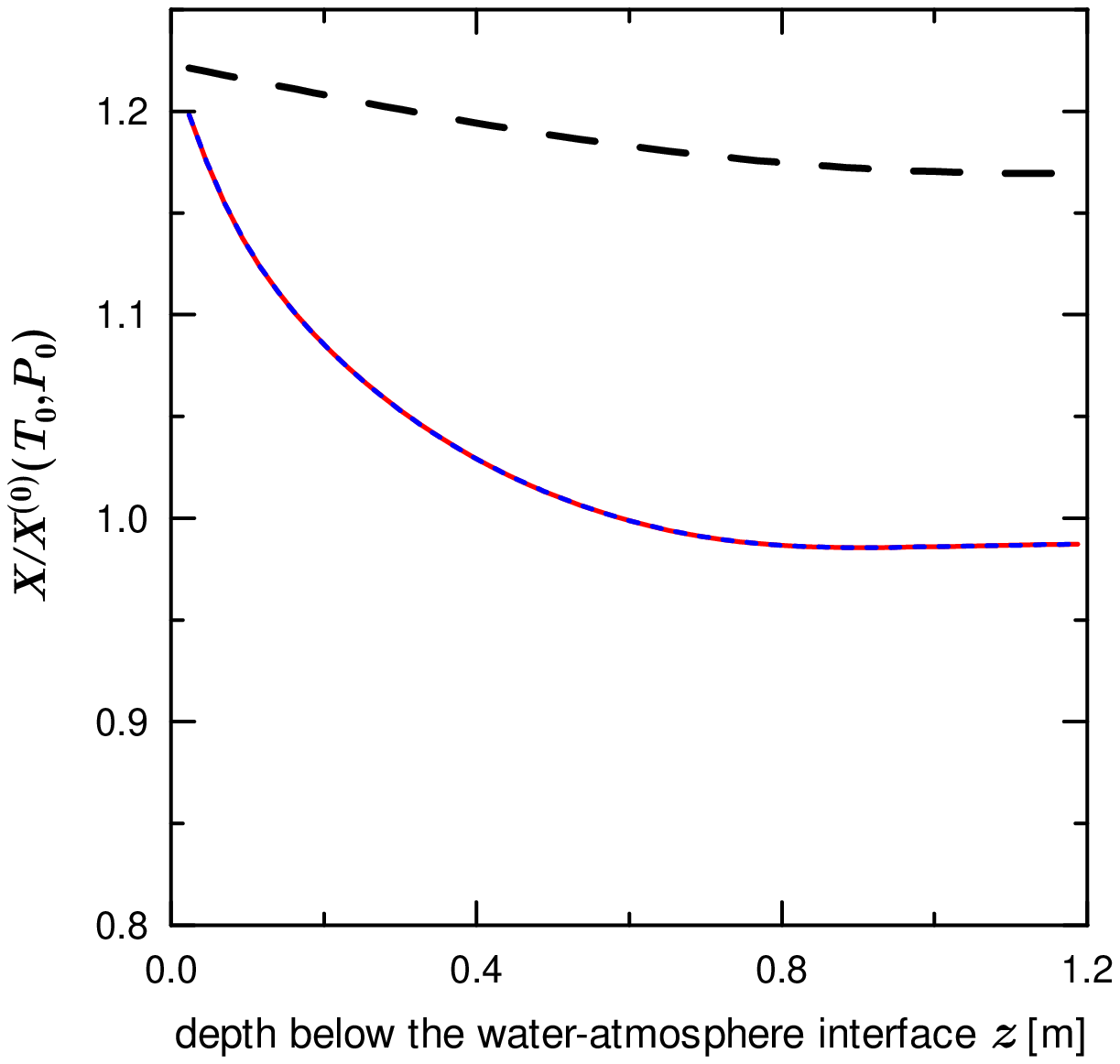}
\qquad\qquad
{\sf (d)}\hspace{-12pt}
\includegraphics[width=0.32\textwidth]%
 {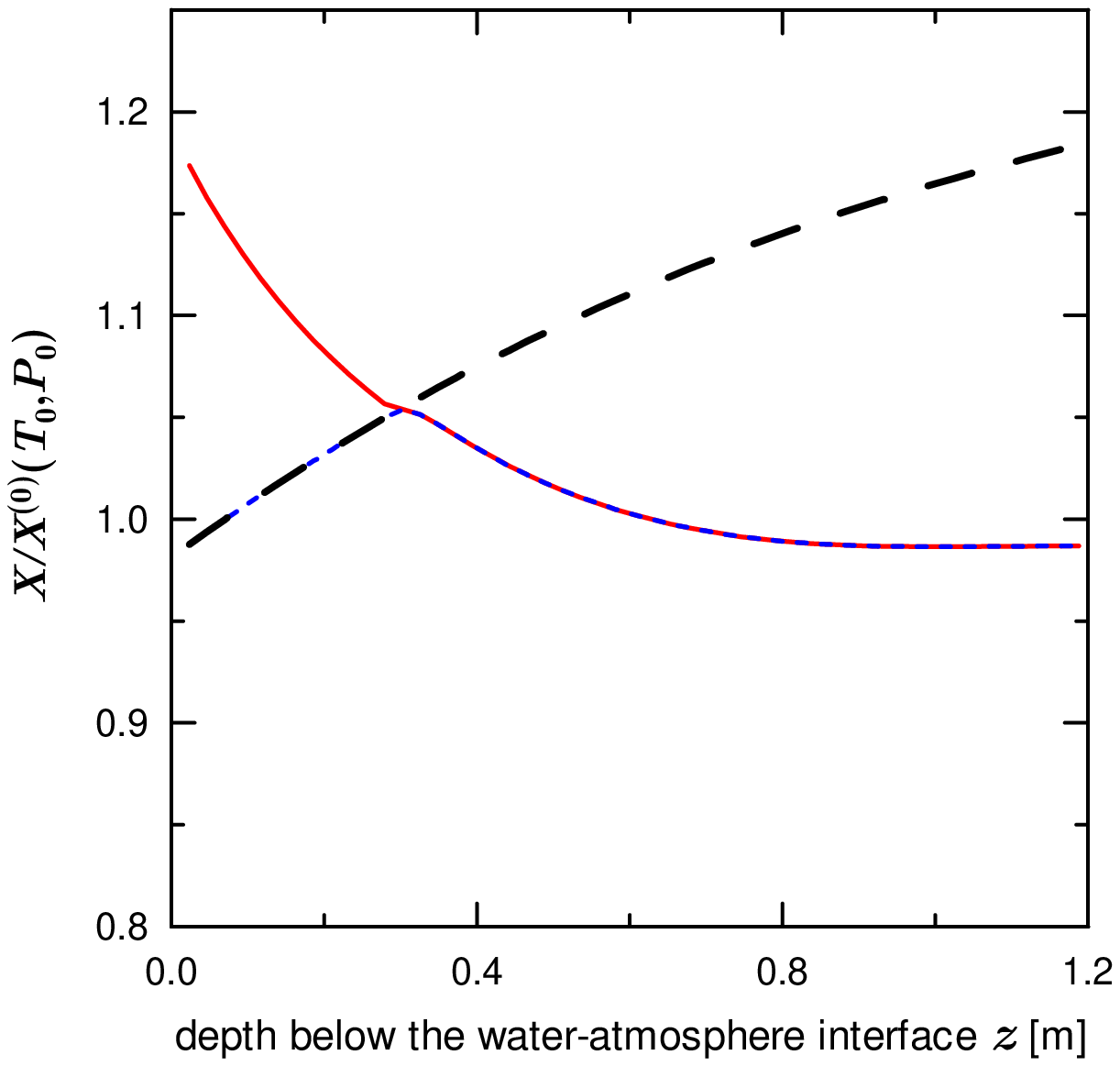}
}

  \caption{(Color online) Snapshots of the oscillating solubility profile $X^{(0)}$ of nitrogen for the annual temperature wave in water-saturated ground (wetland) are plotted with the black dashed lines for different surface temperature oscillation phases: (a)~mid winter, (b)~mid spring, (c)~mid summer, (d)~mid autumn. For simplicity, the atmosphere is assumed to be pure nitrogen, molar fraction of which in the Earth's atmosphere is 78.09\%. The blue dotted lines represent the solution molar concentration $X_s$. The red solid lines show the net molar fraction $X_\Sigma$ of nitrogen molecules in the pore fluid. The bubbly fraction $X_b$ is given by the difference between the red solid and blue dotted profiles (it does not exist for the cold winter period, when solubility is high). Noticeably, the net molar fraction is nearly unchanging during the year.
 }
  \label{fig1}
\end{figure*}
%%%%%%%%%%%%%%%%%%%%%%%%%%%%%%%%%%%%%%%%%%%%%%%%%%%%%%%%%

\subsection{Numerical results}
\label{sec22}
Numerical simulation was performed for the atmosphere composed solely by nitrogen---the single-component atmosphere approximation. Technical details on the numerical simulation are provided in Appendix~\ref{sec_app}. The simulation reveals, that for any initial condition, after a transient process, the system reaches a single stable time-periodic regime. Let us consider this regime.

The linear growth of the solubility with depth owned by the hydrostatic pressure gradient is modulated by the temperature wave~(\ref{eq-21-03}). The oscillating solubility profile (\ref{eq-21-01}) for the temperature wave (\ref{eq-21-03}) and pressure (\ref{eq-21-04}) can be seen in Fig.\,\ref{fig1}. Oscillations of the solubility profile lead to formation of a nearly constant in time profile of the net amount of gas molecules $X_\Sigma(z)$. The profile of net molar fraction $X_\Sigma(z)$ nearly attains the maximal (mid winter) solubility next to the surface, $z=0$; there the bubbly fraction exists for nearly entire year except a short coldest period. Further, $X_\Sigma(z)$ monotonically decreases with depth, along with the decrease of the time interval when the bubbly phase is present, down to the depth where the bubbly phase never appears. Below the latter depth $X_\Sigma(z)=X_s(z)$ is nearly uniform and only slightly changes during the year. Non-uniformity of the profile $X_\Sigma(z)$ in this zone rapidly decays with depth. The asymptotic value $X_\infty$ is close to the annual-mean surface solubility of the gas. These features of the regime can be well seen in Figs.\,\ref{fig2} and \ref{fig3}.

The net molar fraction profile is nearly constant during the year because the molar diffusivity $D$ is 3 orders of magnitude smaller than the heat diffusivity $\chi$, meaning that diffusive redistribution of mass is a slow process against the background of a rapid temperature (and, hence, solubility) oscillation. This well pronounced separation of time scales provides opportunity for developing an analytical theory of the process, allowing for a better insight into the mechanisms of the formation of the bubbly horizon. The results of numerical simulation can be as well more comprehensively understood in the context of this theory.

\subsection{Analytical theory}
\label{sec23}
The principal assumption of our analytical theory is that the net molar fraction profile $X_\Sigma$ is `frozen' for one oscillation period and the period-mean diffusion flux performs a slow diffusive transfer of solute mass. For the analytical treatment we also linearise temperature dependencies of  diffusion coefficient
\begin{equation}
D\approx D_0(1+\delta_1e^{-kz}\cos(\omega t-kz))
\label{eq-23-01}
\end{equation}
and solubility
\begin{equation}
X^{(0)}\approx X_0^{(0)}(1+bz)(1-a_1e^{-kz}\cos(\omega t-kz))\,,
\label{eq-23-02}
\end{equation}
where $a_1=-(\partial\ln{X^{(0)}}/\partial T)_P\,\Theta_0$, $\delta_1=(\partial\ln{D}/\partial T)_P\,\Theta_0$, $b=\rho g/P_0$.

Prior to constructing the analytical theory, let us emphasize, that this analytical theory is an approximation but not a limiting case. The lenearisations (\ref{eq-23-01}) and (\ref{eq-23-02}) require small $\Theta_0$. Meanwhile, for small $\Theta_0$ the penetration depth of the bubbly zone is small (below in the text it will be shown to be nearly linear function of $\Theta_0$) and can become commensurable with the thickness of the diffusion boundary layer $\delta_\mathrm{diff}=\sqrt{2D/\omega}$. In the latter case the approximation of `frozen' profile $X_\Sigma(z)$ is invalid. Thus, the frozen profile approximation is not compatible with the limit of vanishing $\Theta_0$. Nonetheless, for moderate $\Theta_0$, both approximations are satisfactory accurate.

%%%%%%%%%%%%%%%%%%%%%%%%%%%%%%%%%%%%%%%%%%%%%%%%%%%%%%%%%
\begin{figure*}[!t]
\centerline{
{\sf (a)}\hspace{-15pt}
\includegraphics[width=0.31\textwidth]%
 {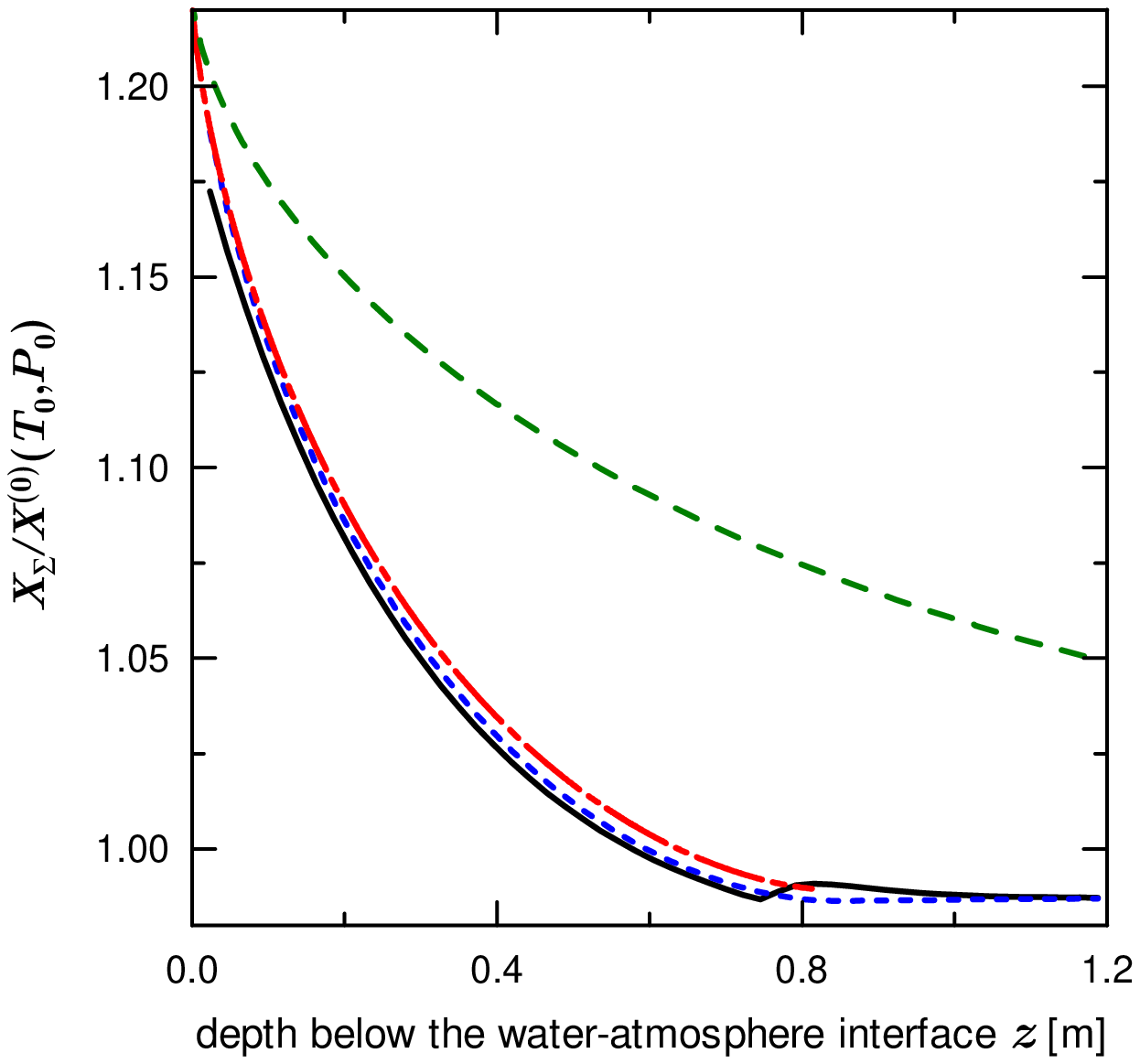}
\quad
{\sf (b)}\hspace{-15pt}
\includegraphics[width=0.31\textwidth]%
 {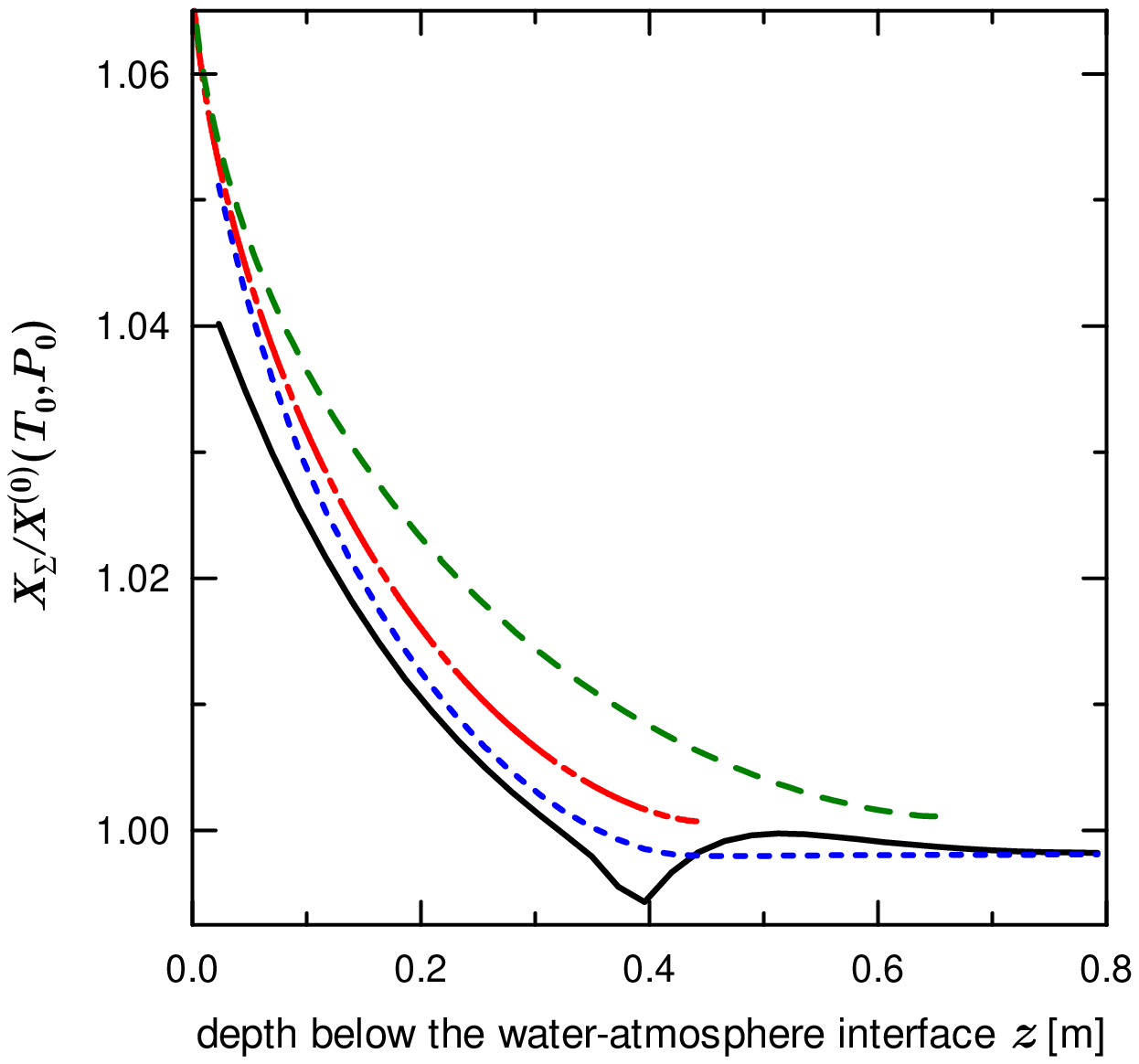}
\quad
{\sf (c)}\hspace{-15pt}
\includegraphics[width=0.31\textwidth]%
 {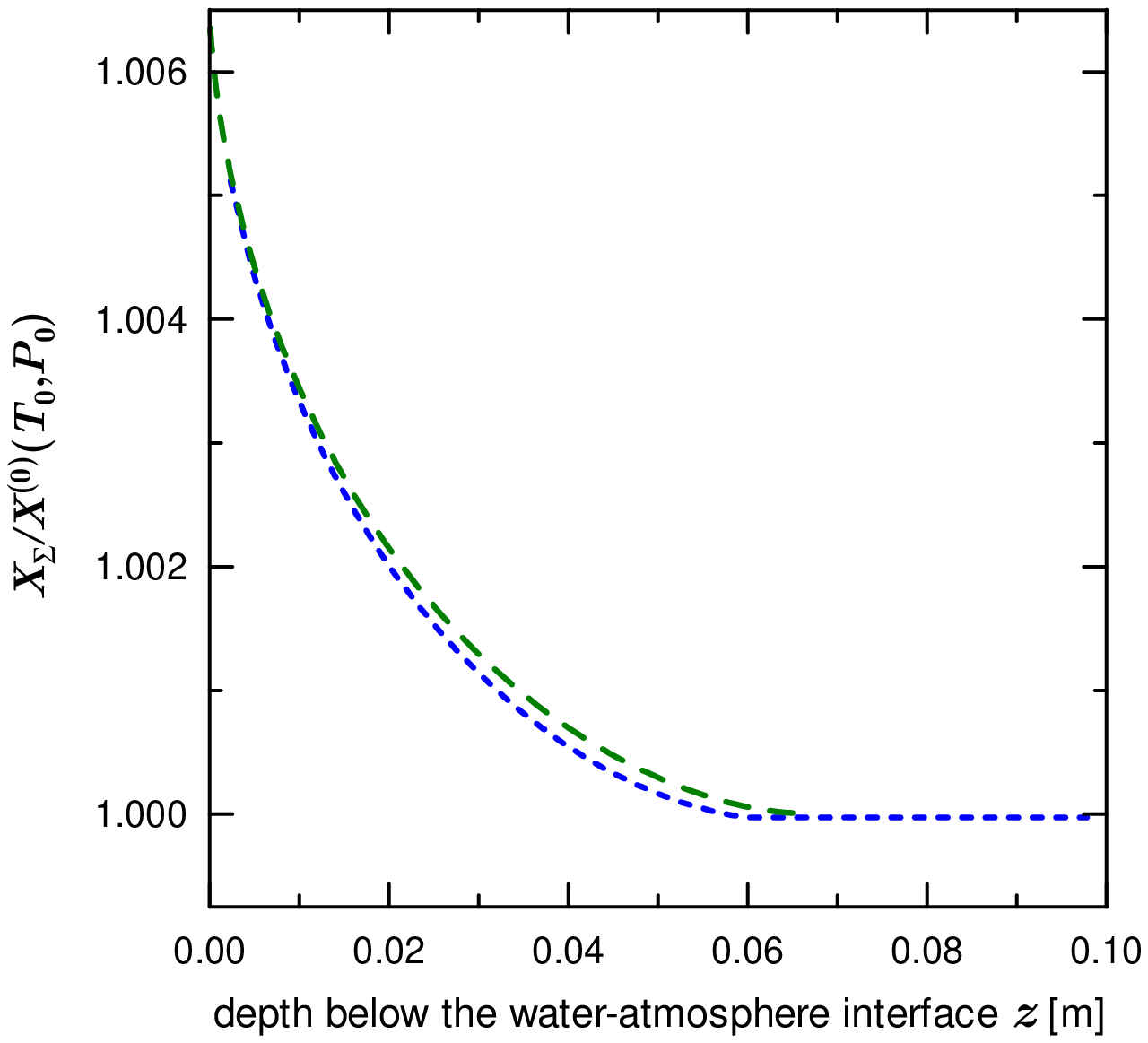}
}

  \caption{(Color online) Profiles of the net molar fraction $X_\Sigma$ are plotted for temperature oscillation amplitudes $\Theta_0=15\,\mathrm{K}$ (a), $\Theta_0=5\,\mathrm{K}$ (b) and $\Theta_0=0.5\,\mathrm{K}$ (c). The black solid lines represent the results of numerical simulation for the real molecular diffusion coefficient of nitrogen in water; the blue dotted lines represent the results of numerical simulation with diminished values of the diffusion coefficient: $D=0.01D_\mathrm{N_2,H_2O}$ in (a) and (b), $D=10^{-4}D_\mathrm{N_2,H_2O}$ in (c). The red dash-dotted lines represent the analytical solution~(\ref{eq-23-07}); the green dashed lines represent the limiting case analytical solution~(\ref{eq-23-09}).
 }
  \label{fig2}
\end{figure*}
%%%%%%%%%%%%%%%%%%%%%%%%%%%%%%%%%%%%%%%%%%%%%%%%%%%%%%%%%

For analytical calculations we take in account the order of magnitude of the relevant system parameters; $a_1|_{\Theta_0=15\,\mathrm{K}}\approx0.2$, $\delta_1\sim a_1$, $b=0.1\,\mathrm{m^{-1}}$, $k\approx 0.8\,\mathrm{m^{-1}}$ for the annual temperature oscillation in wetlands.

At a steady regime the annual-mean diffusive flux $\langle{J}\rangle$ is zero. For the porous medium domain where the bubbly phase appears for some part of the oscillation period, the mean flux reads
\begin{equation}
\langle{J}\rangle =\frac{1}{t_p}\int\limits_{t_1}^{t_2}\left(-D\frac{dX^{(0)}}{dz}\right)dt +\frac{1}{t_p}\int\limits_{t_2}^{t_1+t_p}\left(-D\frac{dX_\Sigma}{dz}\right)dt\,,
\label{eq-23-03}
\end{equation}
where $t_p=2\pi/\omega$ is the oscillation period, $t_1<t_2$ are the time instants between which local temperature is high enough so that the local solubility becomes smaller than $X_\Sigma$. Eq.\,(\ref{eq-23-03}) can be rewritten in terms of the temperature oscillation phase $\varphi=\omega t-kz$;
\begin{eqnarray}
\langle{J}\rangle =\frac{1}{2\pi}\int\limits_{-\varphi_\ast}^{\varphi_\ast}\left(-D\frac{dX^{(0)}}{dz}\right)d\varphi \qquad\nonumber\\
{}+\frac{1}{2\pi}\int\limits_{\varphi_\ast}^{\varphi_\ast+2\pi}\left(-D\frac{dX_\Sigma}{dz}\right)d\varphi
\label{eq-23-04}
\end{eqnarray}
with $\varphi_\ast\in(0,\pi)$ determined by the condition $X^{(0)}(\varphi=\varphi_\ast)=X_\Sigma$, {\it i.e.},
\begin{equation}
1-a_1e^{-kz}\cos\varphi_\ast=\frac{1}{1+bz}\frac{X_\Sigma}{X_0^{(0)}}\,.
\label{eq-23-05}
\end{equation}

After laborious but straightforward mathematical manipulations, to the leading order, equation $\langle{J}\rangle=0$ with substitutions (\ref{eq-23-01}), (\ref{eq-23-02}), (\ref{eq-23-04}) and (\ref{eq-23-05}) takes the form
\begin{eqnarray}
b\,\varphi_\ast+a_1(-b+k+\delta_1b)e^{-kz}\sin\varphi_\ast\qquad\qquad
\nonumber\\[7pt]
 {}+a_1k\delta_1\Big(\varphi_\ast-\frac{1}{2}\sin 2\varphi_\ast\Big)\quad
\nonumber\\[7pt]
 {}+(\pi-\varphi_\ast-a_1\delta_1e^{-kz}\sin\varphi_\ast)\qquad\qquad
\nonumber\\[7pt]
 \times\Big(b-\frac{d}{dz}(a_1e^{-kz}\cos\varphi_\ast)\Big)=0\,.
\label{eq-23-06}
\end{eqnarray}
The latter equation should be treated as an initial value problem; it should be integrated from $\varphi_\ast(z=0)=\pi$ till $\varphi_\ast(z)$ attains $0$, the point $z_b$ where $\varphi_\ast=0$ is the base of the bubbly zone. For $z>z_b$, the bubbly phase never appears and $X_\Sigma$ is constant.
More convenient is to deal with Eq.\,(\ref{eq-23-06}) in terms of $\xi=(b/a_1)z$, $\varkappa\xi=kz$ and $F=e^{-kz}\cos\varphi_\ast$;
\begin{eqnarray}
\frac{d\xi}{dF}=
\frac{\qquad\qquad \pi-\arccos(Fe^{\varkappa\xi}) \ {}}
 {\pi+\varkappa\delta_1e^{-2\varkappa\xi}\arccos(Fe^{\varkappa\xi}) \ {}}
\qquad\qquad\quad
\nonumber\\[15pt]
\frac{{}-a_1\delta_1\sqrt{e^{-2\varkappa\xi}-F^2}\qquad\qquad\quad}
 {{}+\left(\frac{a_1(k-b)}{b}-\frac{\varkappa\delta_1}{4}F\right)\sqrt{e^{-2\varkappa\xi}-F^2}}\,.
\label{eq-23-07}
\end{eqnarray}
The latter equation should be integrated from $F=-1$, $\xi=0$, which corresponds to the surface, till the equality $F=e^{-\varkappa\xi}$ is fulfilled, which corresponds to the base of the bubbly zone. In Figs.\,\ref{fig2}a and b the result of integration of Eq.\,(\ref{eq-23-07}) is plotted with the red dash-dotted lines. One can see that the analytical theory is in a good agreement with the results of numerical simulation.

Eq.\,(\ref{eq-23-06}) can be simplified for the limiting case $a_1\to0$ (where the depth of penetration of the bubbly zone is also small, meaning $kz\to0$);
\begin{equation}
\varphi_\ast+(\pi-\varphi_\ast)\left(1+\sin\varphi_\ast\frac{d\varphi_\ast}{d\xi}\right)=0\,.
\label{eq-23-08}
\end{equation}
The latter equation can be integrated and yields
\begin{equation}
(\pi-\varphi_\ast)\cos\varphi_\ast+\sin\varphi_\ast=\pi\xi\,.
\label{eq-23-09}
\end{equation}
For solution (\ref{eq-23-09}), coordinate $\xi$ varies from $0$ (surface) to $\xi_b=1$ (the base of the bubbly zone). The net molar fraction profile is given by $X_\Sigma/X_0^{(0)}=1+a_1\xi-a_1\cos\varphi_\ast$. For this limiting solution, one can determine the scaling properties of the solution next to the surface, for $\xi\ll 1$;
\[
\frac{X_\Sigma}{X_0^{(0)}}=1+a_1
 -\frac{a_1}{2}\left(\frac{3}{2}\pi\right)^\frac{2}{3}\xi^\frac{2}{3}
 +\dots\,.
\]
This limiting solution is plotted in Fig.\,\ref{fig2} with the green dashed lines. Unfortunately, this limiting solution is accurate only for small amplitudes of temperature oscillations, where the diffusive boundary layer $\xi_\mathrm{diff}=(b/a_1)(2D/\omega)^{1/2}$ is commensurable with the bubbly zone thickness $\xi_b=1$ and thus the assumption of the `frozen' profile is not valid for typical diffusivities in liquids.

In order to ensure the correctness of analytical calculations, numerical simulation was performed for a diminished molecular diffusion $D$. The molecular diffusion coefficient was small enough for both assumptions of the analytical theory can be simultaneously accurate. In Figs.\,\ref{fig2}a and b, one can see that the results of analytical theory (\ref{eq-23-07}) match the results of numerical simulation for $100$-fold diminished $D$ plotted with blue dotted lines. In Fig.\,\ref{fig2}c, the limiting analytical solution (\ref{eq-23-09}) matches the numerical results for the diffusion coefficient diminished by factor $10^4$. Considering results of numerical simulation for diminished diffusivity, one can notice disappearance of the kink near the boundary between the bubbly horizon and the zone of undersaturated solution, which can be seen in Figs.\,\ref{fig1} and \ref{fig2}a,b for `normal' diffusion strength. Thus, one can conclude that this kink is related to finiteness of the ratio $D/\chi$. The $X_\Sigma$-profile is not completely frozen on the time scale of temperature oscillations; at the point of discontinuity of the solute concentration gradient even a small diffusivity can result in an observable distortion of the net profile $X_\Sigma$.

The case of impaired diffusion can be also of physical interest for the porous media where the pore network is not globally connected and diffusive transport necessarily involves diffusion through the solid matrix material separating different connected clusters of pores from each other. In this case the effective diffusion coefficient will be diminished by several orders of magnitude compared to the bulk diffusivity in the pore liquid.

%%%%%%%%%%%%%%%%%%%%%%%%%%%%%%%%%%%%%%%%%%%%%%%%%%%%%%%%%
\begin{figure*}[!t]
\centerline{
{\sf (a)}\hspace{-12pt}
\includegraphics[width=0.33\textwidth]%
 {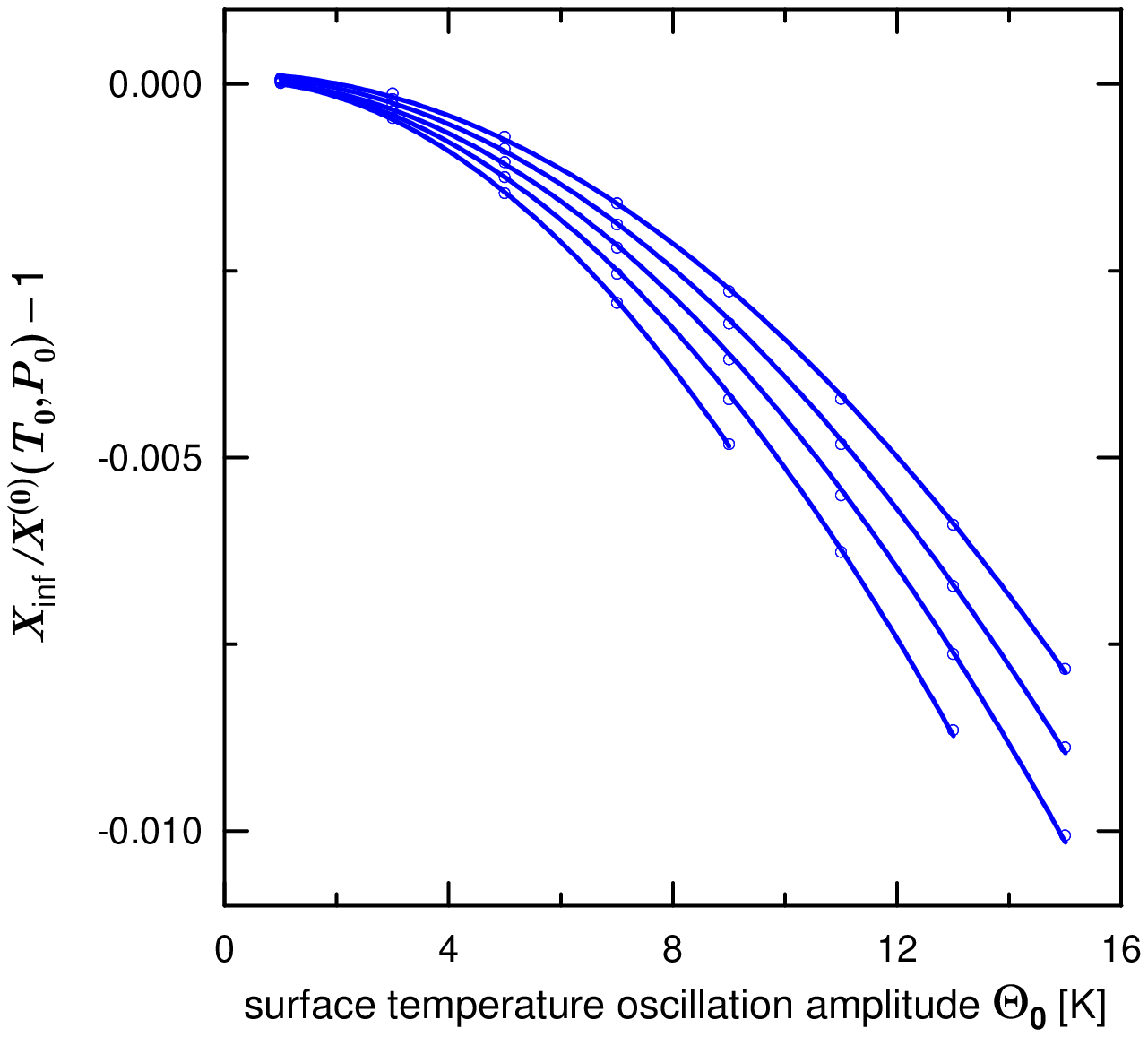}
\qquad\qquad
{\sf (b)}\hspace{-12pt}
\includegraphics[width=0.33\textwidth]%
 {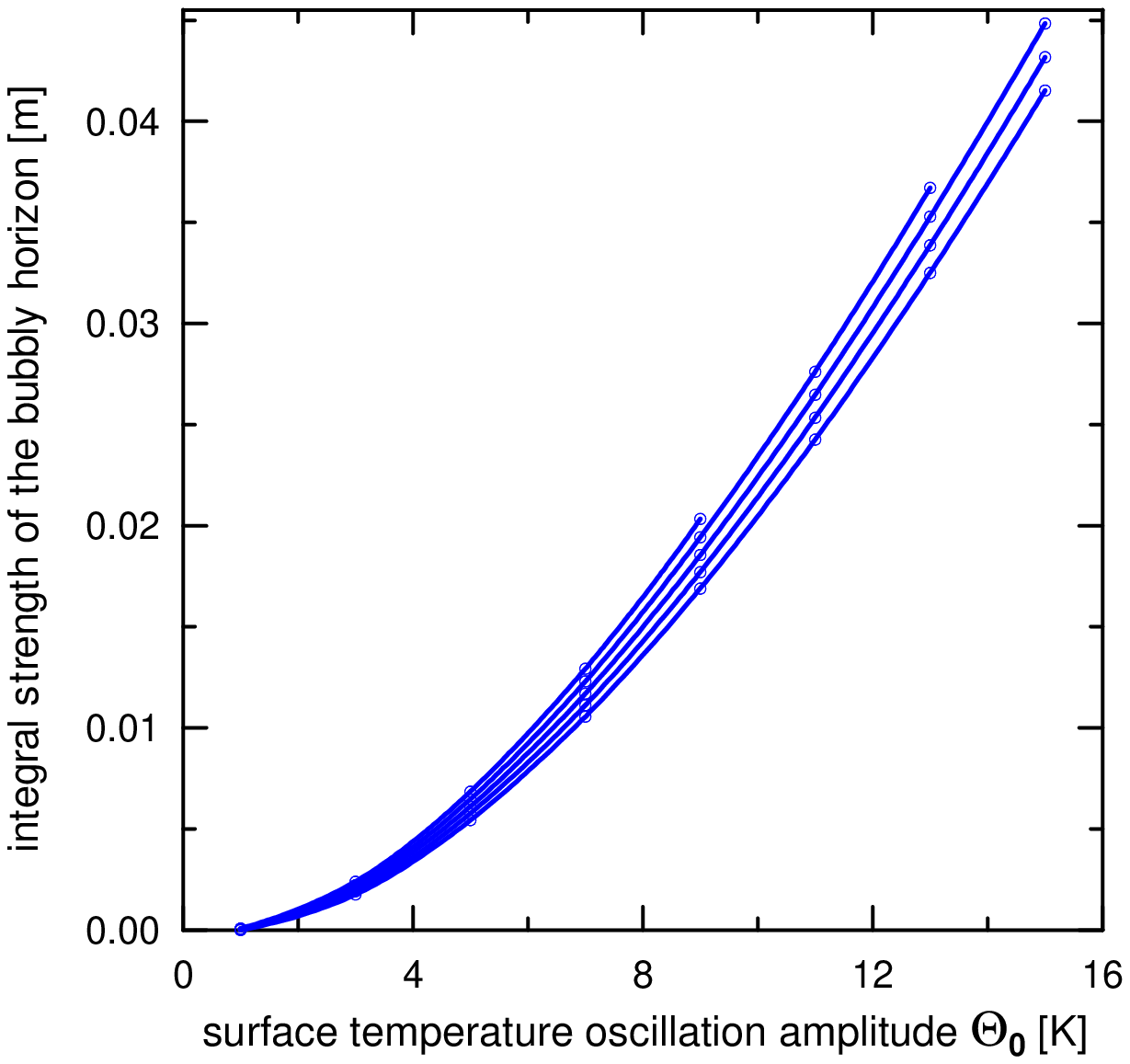}
}

  \caption{(Color online) For nitrogen atmosphere and water-saturated porous medium the bubbly horizon quantifiers are plotted {\it vs} the surface temperature oscillation amplitude $\Theta_0$ for annual-mean temperature $T_0=9$, $13$, $17$, $21$, $25\,\mathrm{K}$ (from bottom to top in (a) and from top to bottom in (b)). In panel (a) the lines are a quadratic fitting of the calculated data plotted with symbols.
 }
  \label{fig3}
\end{figure*}
%%%%%%%%%%%%%%%%%%%%%%%%%%%%%%%%%%%%%%%%%%%%%%%%%%%%%%%%%

\subsection{Integral quantifiers of the bubbly horizon}
\label{sec24}
The developed analytical theory suggests natural quantifiers of the $X_\Sigma$-profile: the deep solute concentration $X_\infty/X_0^{(0)}$ and the integral strength of the bubbly layer $I=\langle\int[(X_\Sigma-X_\infty)/X_0^{(0)}]dz\rangle$. These characteristics are good quantifiers of the system state even for situations where the $X_\Sigma$-profile is far from being frozen (see Fig.\,\ref{fig2}). In Fig.\,\ref{fig3}, quantifiers $X_\infty$ and $I$ are plotted for nitrogen atmosphere and water-saturated porous medium. The deviation of $X_\infty$ from the annual-mean near-surface solubility is a nonlinear effect, quadratic in amplitude $\Theta_0$. For non-vanishing $\Theta_0$ the effect is negative, {\it i.e.}, the temperature oscillation results in `ventilation' of deep areas of the porous medium.

%%%%%%%%%%%%%%%%%%%%%%%%%%%%%%%%%%%%%%%%%%%%%%%%%%%%%%%%%
\begin{figure*}[!t]
\centerline{
{\sf (a)}\hspace{-12pt}
\includegraphics[width=0.32\textwidth]%
 {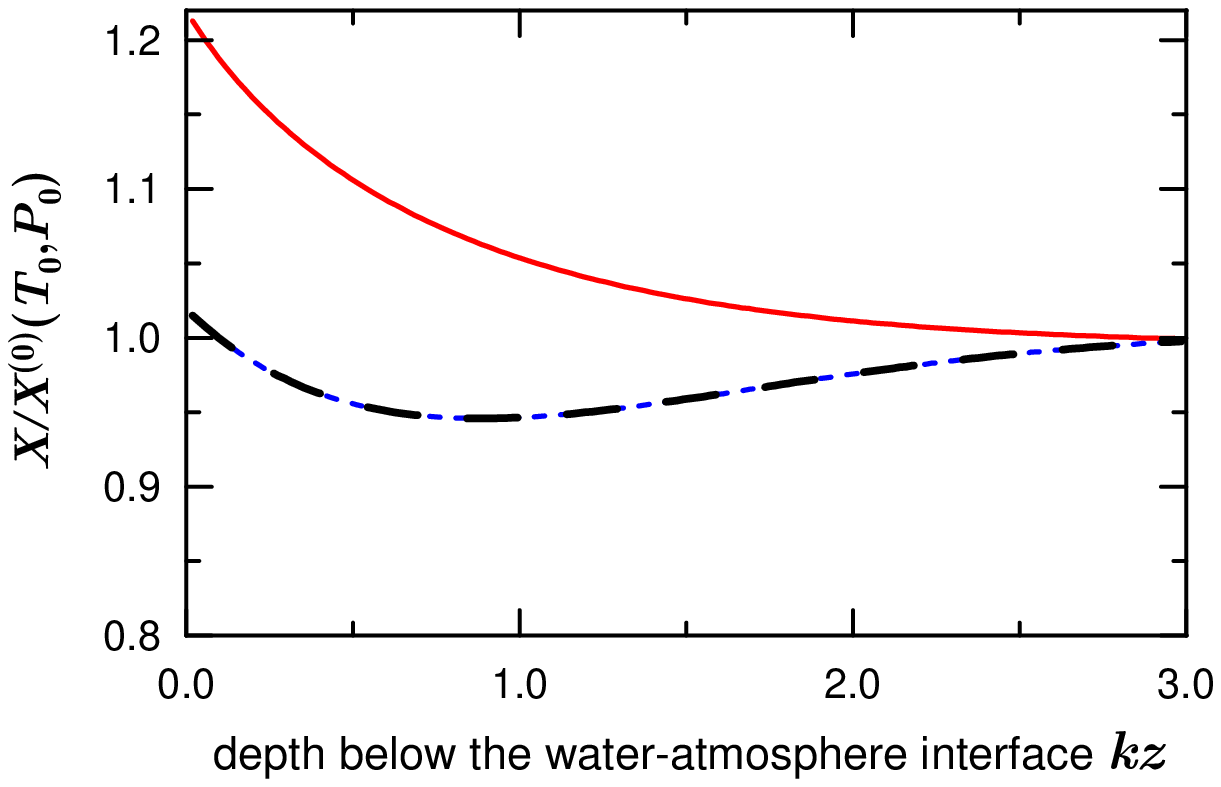}
\qquad\qquad
{\sf (b)}\hspace{-12pt}
\includegraphics[width=0.32\textwidth]%
 {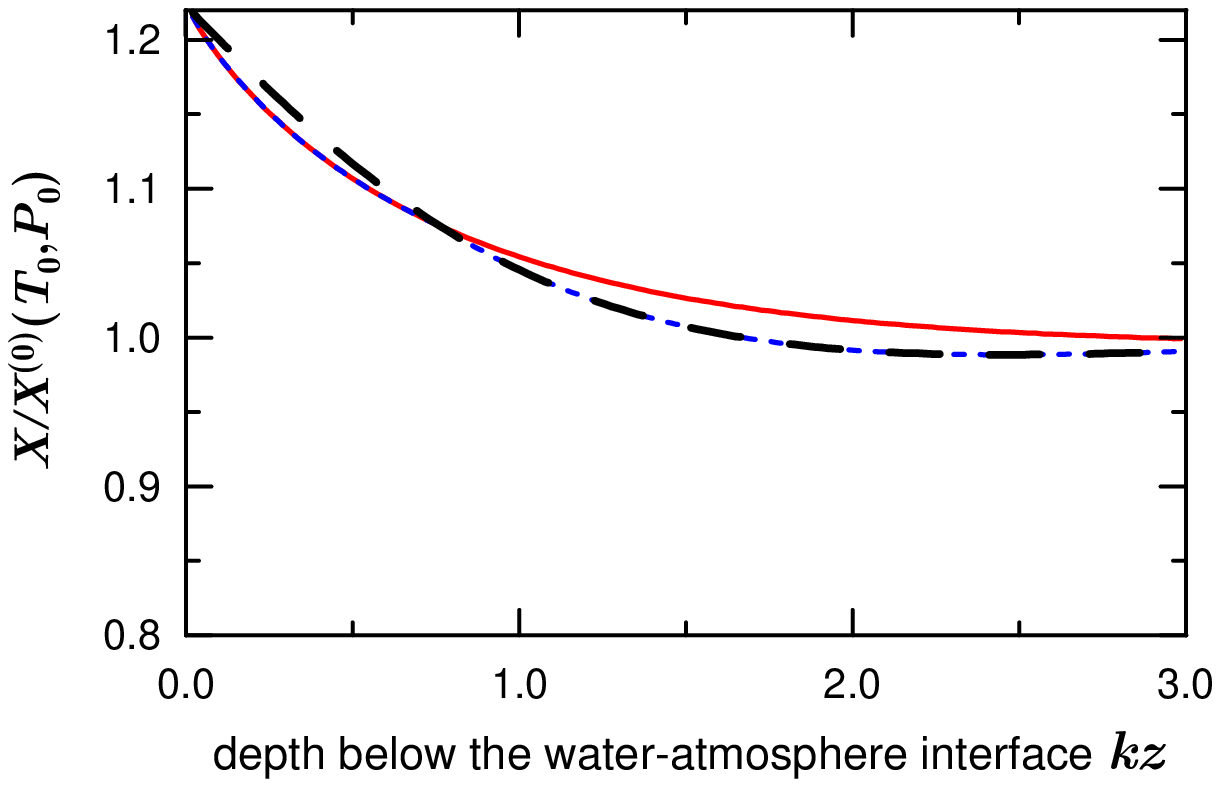}
}
\vspace{20pt}

\centerline{
{\sf (c)}\hspace{-12pt}
\includegraphics[width=0.32\textwidth]%
 {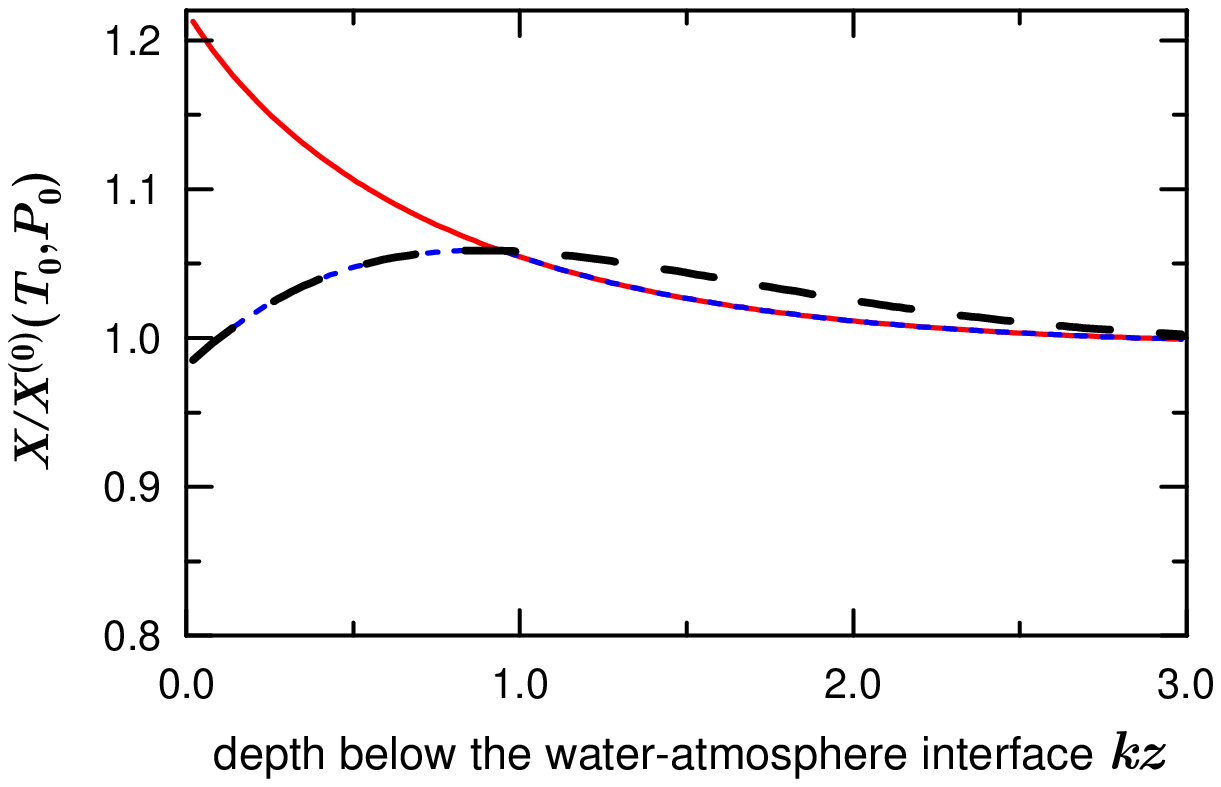}
\qquad\qquad
{\sf (d)}\hspace{-12pt}
\includegraphics[width=0.32\textwidth]%
 {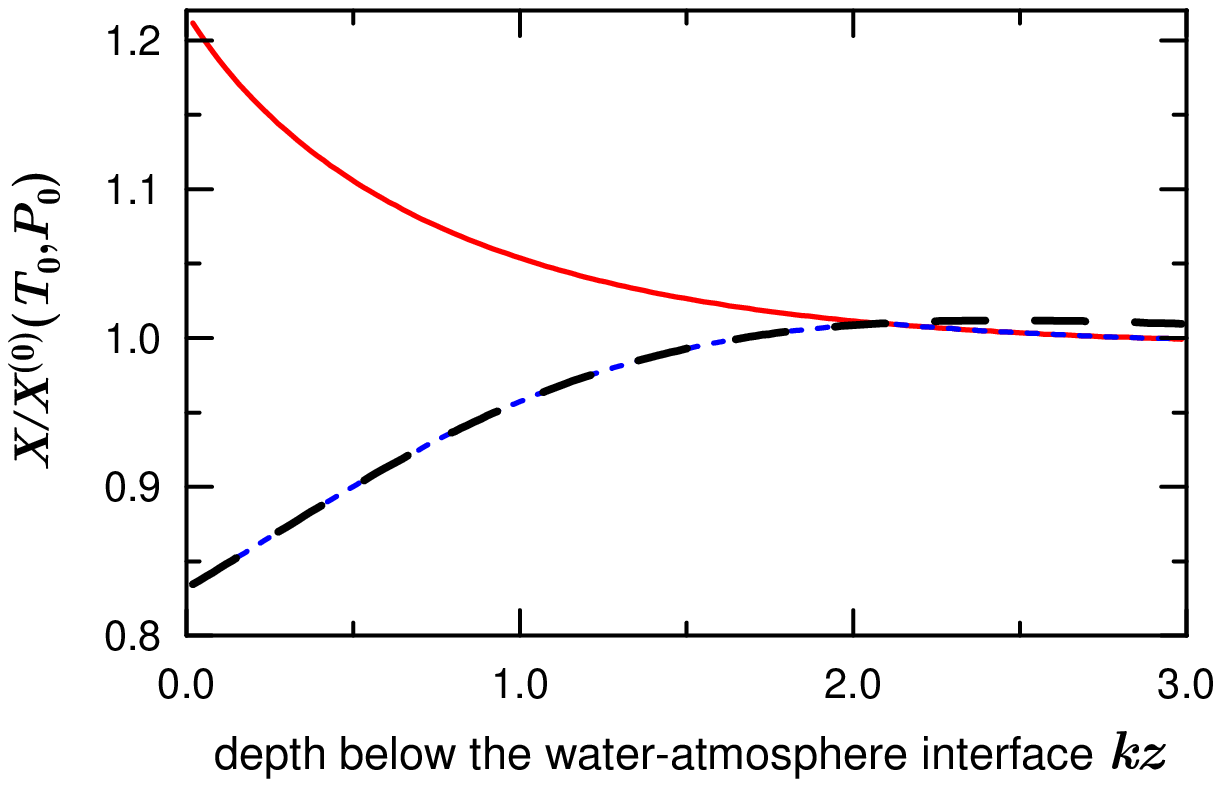}
}

  \caption{(Color online) Dynamics of the bubbly horizon in porous medium in the absence of the hydrostatic pressure gradient is presented for the nitrogen atmosphere. Snapshots of profiles of $X^{(0)}(z)$ (black dashed lines), $X_s(z)$ (blue dotted lines) and $X_\Sigma(z)$ (red solid lines) are plotted for different phases $\varphi$ of the surface temperature oscillation: (a)~$\varphi(z=0)=0$, (b)~$\pi/2$, (c)~$\pi$, (d)~$3\pi/2$. The results are provided in the dimensionless form and are valid for arbitrary oscillation frequency.
 }
  \label{fig4}
\end{figure*}
%%%%%%%%%%%%%%%%%%%%%%%%%%%%%%%%%%%%%%%%%%%%%%%%%%%%%%%%%

\section{The case of no effect of pressure gradient}
\label{sec3}
For the annual temperature wave the penetration depth $1/k=\sqrt{2\chi/\omega}$ is about $1\,\mathrm{m}$; for this depth the hydrostatic pressure increase is comparable to the atmospheric pressure and, thus, the gas solubility is significantly influenced by the hydrostatic pressure gradient. For higher oscillation frequencies in nature ({\it e.g.}, daily oscillations) and technological systems the hydrostatic pressure increase for the wave penetration depth can be negligible. The case of solutions of solids and liquids is qualitatively similar to the case of no hydrostatic pressure gradient, as the solubility of condensed matter is nearly independent of pressure (see Eq.\,(\ref{eq-21-02})).

In Fig.\,\ref{fig4}, one can see that in the absence of pressure gradient the bubbly horizon is not bounded from below. For any depth $z$ the solubility profile $X^{(0)}$ exceeds the $X_\Sigma$-profile ({i.e.}, the bubbly phase locally disappears) for a small fraction of the oscillation period; $X_\Sigma(z)$ is slightly smaller than the envelope of the wave of $X^{(0)}(z,t)$. For the limit of vanishingly small ratio $D/\chi$ the $X_\Sigma$-profile coincides with the envelope of $X^{(0)}(z,t)$. While the envelopes of solubility profiles are different for gaseous and condensed matter, the principle, that $X_\Sigma(z)$-profile tends to coincide with the envelope of the solubility profile wave, remains valid for solids and liquids. Hence, the integral strength of the bubbly horizon approximately is
\begin{equation}
I=\int_0^{+\infty}a_1e^{-kz}dz=\left|\frac{\partial\ln X^{(0)}}{\partial T}\right|_P\frac{\Theta_0}{k}\,.
\label{eq-31-01}
\end{equation}

\section{Conclusion}
\label{sec4}
We have considered the diffusion transport of a weakly soluble substance in a liquid-saturated porous medium half-space being in contact with the reservoir of this substance in the case where the surface temperature harmonically oscillates in time. The surface temperature oscillation creates the temperature wave which propagates deep into porous medium and decays with depth. The solubility wave associated with the temperature wave results in time-dependent intermittency of the zones of non-dissolved phase with saturated solution and the zones of undersaturated solution.

Due to the smallness of the ratio $D/\chi$, which is as small as $\sim10^{-3}$ for transport processes in liquids, the diffusion transport in the system is much slower than the temperature (and solubility) variation. As a result, the profile of the net molar fraction of `guest' molecules in pores, $X_\Sigma$ (`net' means `solute + non-dissolved phase'), remains nearly constant over the oscillation period. We have revealed {\em for gases} that the $X_\Sigma$-profile nearly attains the maximal-per-period solubility next to the surface and monotonously decays with depth in the zone where the non-dissolved phase can be observed, the bubbly horizon, and reaches a constant level beneath the bubbly horizon (Fig.\,\ref{fig2}). The bottom boundary of the bubbly horizon is controlled by the hydrostatic pressure gradient.

Without the pressure gradient, the bubbly horizon is not bounded from below and the $X_\Sigma$-profile is slightly smaller than the envelope of the oscillating solubility profile (Fig.\,\ref{fig4}). At any depth $z$ the bubbly phase disappears only for a short part of the oscillation period; in the idealistic limiting case $D/\chi\to0$ this part of oscillation period tends to $0$. Notice, the solubility profile decays with depth exponentially and, thus, even though the bubbly horizon is not formally bounded from below, it becomes vanishingly weak at the depth. The no-pressure-gradient case is relevant for gases and short-term oscillations, when the penetration depth of the temperature wave is small compered to the depth scale of a noticeable increase of the hydrostatic pressure, and for condensed phases, solubility of which is insensitive to pressure.

For the reported results an analytical theory, which lends better insight into the mechanisms of the phenomenon, has been developed [Eqs.\,(\ref{eq-23-07}) and (\ref{eq-23-09})].

For the annual temperature oscillation of amplitude $15\,\mathrm{K}$ the atmosphere gas bubbly horizon has been found to have $0.8\,\mathrm{m}$ penetration depth and near-surface relative increase of $X_\Sigma$ by $20\%$ compared to the no-oscillation solubility (Fig.\,\ref{fig2}a). Such an effect can be considered non-negligible for the local conditions for microflora and fauna. The dependence of integral quantifiers of the bubbly horizon on average surface temperature and temperature oscillation amplitude is plotted in Fig.\,\ref{fig3}.

The periods of negative temperatures with frozen groundwater are beyond the scope of this study and will be considered elsewhere.

Authors acknowledge financial support by the Government of Perm Region (Contract C-26/0004.3) and the Russian Foundation for Basic Research (project no. 14-01-31380\_mol\_a).

\appendix
\section{Numerical simulation}
\label{sec_app}
The evolution of Eq.\,(\ref{eq-21-07}) was simulated with a finite difference method, explicit scheme with central differences. The time-dependent fields of temperature, molecular diffusion coefficient, and solubility were precalculated for one cycle of the surface temperature oscillation on the space--time grid used for simulation of Eq.\,(\ref{eq-21-07}). The calculation domain was limited from below by the depth $L_\infty$, and boundary condition~(\ref{eq-21-09}) was moved to $z=L_\infty$; the depth $L_\infty$ was chosen so that perturbations of the solute concentration profile were not visually resolvable at $z=L_\infty/2$. The choice of the space stepsize was guided by two requirements. Firstly, the decrease of the stepsize by half should not change the results of simulation more than by $1\%$. Secondly, for the diffusion coefficient diminished by factor $10^{-4}$ the relative mismatch between the results of numerical simulation and unsimplified analytical solution (\ref{eq-23-06}) should not exceed $0.1\%$. Practically, the grids with $30$--$50$ points per the zone of non-dissolved phase were sufficiently detailed for these requirements to be met. The time stepsize was determined by the requirement of stability of the explicit method.

\end{document}